\documentclass{raa_twocolumn}
\usepackage{graphicx,times}
\usepackage{natbib}
\usepackage{amssymb,amsmath}
\bibpunct{(}{)}{;}{a}{}{,}

\newcommand\target{\rm{J108}}
\usepackage[pagebackref=true]{hyperref}

\begin{document}

   \title{First detailed analysis of a relatively deep, low mass-ratio contact binary: ATO J108.6991+27.8306}

 \volnopage{ {\bf 20XX} Vol.\ {\bf X} No. {\bf XX}, 000--000}
   \setcounter{page}{1}

   \author{Shuo Ma\inst{1,2,3}, Jinzhong Liu\inst{1,2}, Yu Zhang\inst{1,2}, Guoliang L$\rm{\ddot{u}}$\inst{4}, Ting Wu\inst{1,2}, Chenyang He\inst{4}
}

\institute{ Xinjiang Astronomical Observatory, Chinese Academy of Sciences, Urumqi, Xinjiang 830011, People's Republic of China; {\it liujinzh@xao.ac.cn}\\ \and University of Chinese Academy of Sciences, Beijing 100049, People's Republic of China\\ \and Department of Astronomy, Xiamen University, Xiamen, Fujian 361005, People’s Republic of China \and School of Physical Science and Technology, Xinjiang University, Urumqi, 830064, People’s Republic of China\\
\vs \no
   {\small Received 20XX Month Day; accepted 20XX Month Day}
}

\abstract{We present the first detailed photometric analysis of ATO J108.6991+27.8306 (hereinafter as $\target$). The short-period close binary $\target$ was observed by the Nanshan 1-m Wide Field Telescope (NOWT) of the Xinjiang Astronomical Observatory. The obtained $BVRI$-band light curves were used to determine the photometric solution by using the 2003 version of the Wilson-Devinney code. $\target$ is a typical deep ($f$ $>$ 50\%), low mass ratio ($q$ $<$ 0.25) (DLMR) overcontact binary system with a mass ratio of $q$ = 0.1501 and a fill-out factor of $f$ = 50.1\%, suggesting that it is in the late evolutionary stage of contact binary systems. We found the target to be a W-type W UMa binary and provided evidence for the presence of starspots on both components. From the temperature-luminosity diagram, the main component is the evolved main sequence star with an evolutionary age of about 7.94 Gyr.
}

   \authorrunning{S. Ma et al. }            
   \titlerunning{Study of a relatively deep, low mass-ratio contact binary}  
   \maketitle

%
\section{Introduction}           
\label{sect:intro}

W Urase Majoris-type binaries are usually contact binary systems in which both components fill the Roche lobe during their main sequence evolutionary phase and share a common envelope \citep{1968ApJ...151.1123L,2006AcA....56..199S}, so that both components have approximate surface temperatures. W UMa binaries, whose component spectral types typically range from F to K, may have magnetic activity, such as  starspots \citep{2021RAA....21..203P,2021AJ....161..221P,2022arXiv220706255M}. In addition, W UMa binaries with more massive components but at lower temperatures are defined as W-type W UMa binaries, and the opposite as A-type \citep{1970VA.....12..217B}. The orbital period of the W UMa system is usually variable. The mass transfer between the two components \citep{2006AJ....132.2260H,2021MNRAS.503..324M} can cause the orbital period to increase or decrease. The magnetic braking effect \citep{1998MNRAS.296..893L,2020MNRAS.491.5717B} causes a loss of orbital angular momentum, which is expressed as a decrease in the orbital period. The magnetic cycle \citep{2005A&A...441.1087B,2018ARep...62..520K} and the third body \citep{2020RAA....20...96S} are responsible for the cyclic variation of the orbital period.

There are deep ($f>50\%$), low mass ratio ($q<0.25$) (DLMR) overcontact binaries \citep{2006Ap&SS.304...25Q,2015AJ....150...69Y} among W Urase Majoris-type binaries, which are at the evolutionary stage of late short-period close binaries.
These binaries are important for the study of stellar astrophysics because they may be the progenitors of single rapidly rotating stars \citep{2020MNRAS.492.2731J}, such as Blue Straggler (BS) stars \citep{2009ApJ...697.1048P,2021RAA....21..223W} and FK Com-type stars \citep{1981ApJ...247L.131B,2020MNRAS.496..295S}.

ATO J108.6991+27.8306 (TIC 155709042, LPSEB3, UCAC4 590-039991, ASASSN-V J071447.79+274950.6) was defined as close binary with an orbital period of 0.393204 days by \citet{2018AJ....156..241H}. Furthermore, $\target$ was first identified as an EW type binaries by All-Sky Automated Survey for Supernovae \citep[ASAS-SN,][]{2019MNRAS.486.1907J} and confirmed by \citet{2020ApJS..249...18C}. At the same time, this target was included in the LAMOST catalog \citep{2020ApJS..249...31Y} and no more studies have been done on $\target$ since then. In this paper, we found that $\target$ is a member of the DLMR and attempted to perform the first public analysis of this interesting and unusual target.

\section{Observations and Data reduction}

$\target$ was observed on Dec. 23, 2020 using the Nanshan One-meter Wide-field Telescope \citep[hereinafter as NOWT]{2020RAA....20..211B} of the Xinjiang Astronomical Observatory. During our observations, a standard Johnson-Cousins $BVRI$ filter was used. Individual observations included 107 in the B filter, 107 in V, 105 in R, 112 in I and a total of 426 CCD images were acquired. These CCD images correspond to exposure times of 70,40,45,50 s for the four bands of $BVRI$, respectively, with the observational accuracy better than 0.006 mag.

\begin{figure}[htpb]
	\centering
	\includegraphics[width=0.48\textwidth]{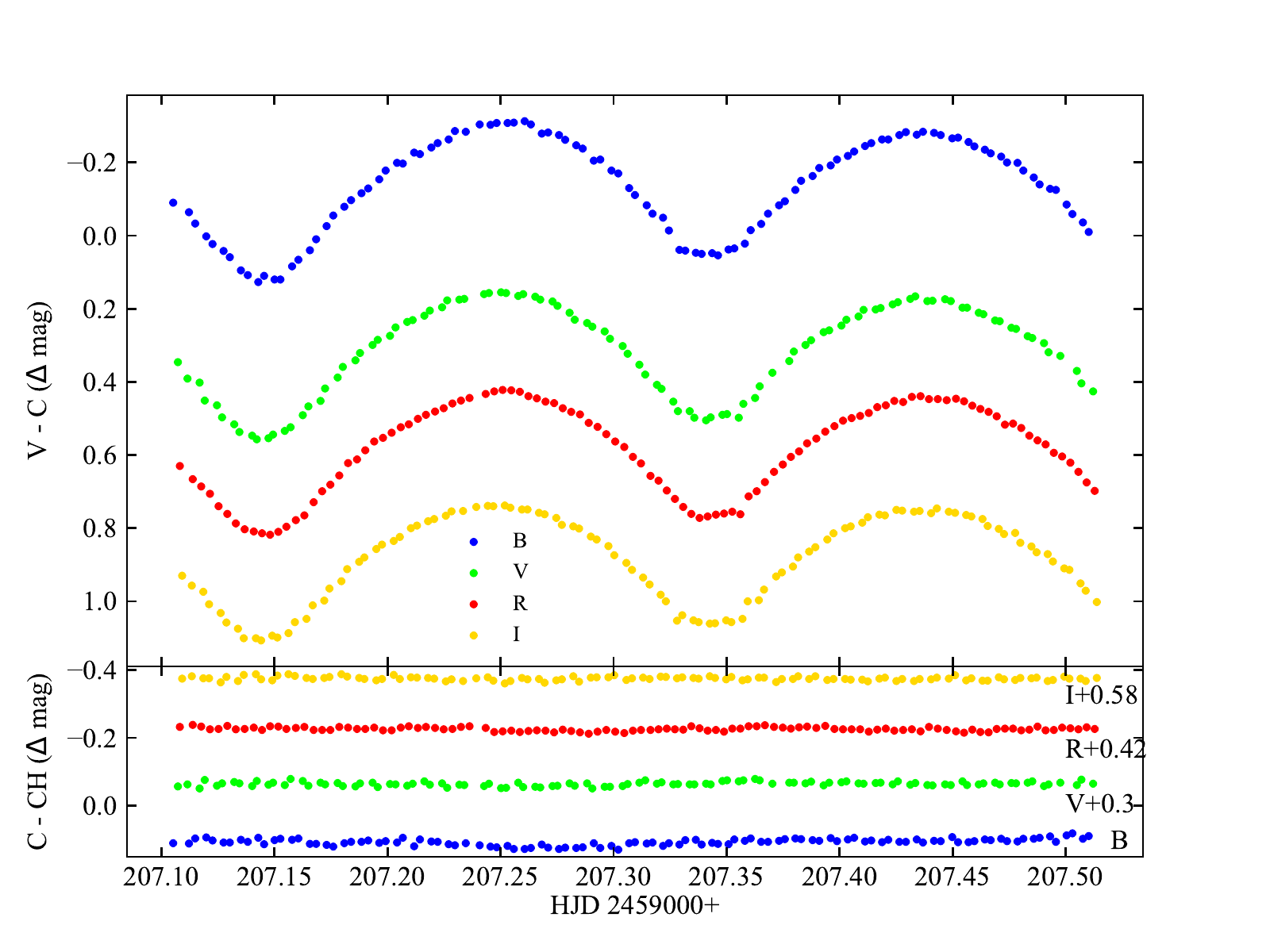}
	\caption{Multicolour light curves of $\target$.}
	\label{fig:hjd}
\end{figure}

The observed CCD images were reduced in a standard manner using IRAF\footnote{the Image Reduction and Analysis Facility \url{http://iraf.noao.edu/}}. In our work, the differential photometry method was adopted. The basic information about variable, comparison, and check stars are compiled in Table \ref{tab:vcch}. The light curves are shown in Figure \ref{fig:hjd}.

\begin{table*}[htp]
	\bc
	\footnotesize
	\caption{Observing details of the variable, comparison and check stars}
	\resizebox{\textwidth}{!}
	{\begin{tabular}{llccccccc}
		\hline\hline
		Targets        & Name                   & $\alpha_{2000}$ & $\beta_{2000}$ & $B\_mag^a$& $V\_mag^a$& $J\_mag^b$&$H\_mag^b$ & $K\_mag^b$ \\ \hline
		Variable star  & ATO J108.6991+27.8306  & 07 14 47.80     & +27 49 50.5    &14.764&14.224& 13.163 & 12.916 & 12.925 \\
		The comparison & 2MASS 07151009+2743415 & 07 15 10.09     & +27 43 41.5    &15.145&13.982&11.922&11.297&11.200\\
		The Check      & 2MASS 07143242+2751055 & 07 14 32.42     & +27 51 05.5    &14.956&14.434&13.330 &13.070&13.022\\ \hline
	\end{tabular}}
	\label{tab:vcch}
	\ec
	\tablecomments{0.98\textwidth}{$^a$The $BV$-band magnitudes of the variable, comparison and check stars were determined from the AAVSO Photometric All-Sky Survey DR9 \citep[APASS9;][]{2016yCat.2336....0H}\\$^b$The $JHK$-band magnitudes were determined from Two Micron All Sky Survey \citep[2MASS;][]{2003yCat.2246....0C}}

\end{table*}

With our observations, the eclipsing times were obtained through the K-W method \citep{1956BAN....12..327K} in Nelson's program\footnote{\url{https://www.variablestarssouth.org/software-by-bob-nelson/}} from the $BVRI$-bands light curves. The new eclipsing times and their mean values are shown in Table \ref{tab:our min}.

\begin{table*}[htp]
	\centering
	\footnotesize
	\caption{Newly Obtained Minima Times of $\target$ in the $BVRI$ Bands}	
	\begin{tabular*}{\hsize}{@{}@{\extracolsep{\fill}}cccccc@{}}
		\hline\hline
		HJD(B)                & HJD(V)                & HJD(R)              & HJD(I)              & HJD(average)           & Min.      \\
		2459000+              & 2459000+              & 2459000+            & 2459000+            & 2459000+               &           \\ \hline
		207.14545$\pm$0.00036 & 207.14488$\pm$0.00043 & 207.14586$\pm$0.00022 & 207.14452$\pm$0.00053 & 207.14518$\pm$0.00039 & primary   \\
		207.34242$\pm$0.00052 & 207.34242$\pm$0.00044 & 207.34296$\pm$0.00035 & 207.34241$\pm$0.00019 & 207.34255$\pm$0.00035 & secondary \\
		\hline
	\end{tabular*}
	\label{tab:our min}
\end{table*}

\section{Photometric analysis} \label{sec:photometric}
To further understand $\target$, we analyzed our $BVRI$-band light curves using the 2013 version of the W-D program \citep{1971ApJ...166..605W,1979ApJ...234.1054W,1990ApJ...356..613W,2008ApJ...672..575W,2012AJ....144...73W,2010ApJ...723.1469W}.
We obtained several $\target$ spectra from LAMOST, as detailed in Table \ref{tab:spect}, where two spectra provide effective temperatures for the binary of 6407.81 $\pm$ 96.88 K and 6267.75 $\pm$ 24.80 K. Combined with Gaia DR2 \citep{2018A&A...616A...1G}, DR3 \citep{2022yCat.1355....0G}, and TESS \citep{2022yCat.4039....0P} provided effective temperatures of 6214 $\pm$ 395 K, 5888.2 $\pm$ 17.1 K, and 6264.9 $\pm$ 18.1 K, respectively, the average value of 6208 $\pm$ 110 K calculated from these temperatures was taken as the effective temperature of the binary. We assumed for the time being that the surface temperature $T_1$ of the primary component is consistent with this effective temperature. Based on this temperature, the gravity-darkening coefficients and the bolometric albedo were set at $ g_{1, 2} = 0.32 $ \citep{1967ZA.....65...89L} and $ A_{1, 2} = 0.5 $ \citep{1969AcA....19..245R} in either binary system, respectively. The bolometric and bandpass limb-darkening were adopted the square-root functions laws from the values of \citet{1993AJ....106.2096V}.

\begin{table*}[htpb]
	\caption{LAMOST spectral information of $\target$}
	\centering
	\footnotesize
	\resizebox{\textwidth}{!}
	{\begin{tabular}{cccccccc}
		\hline\hline
		UT Date&HJD  &phase &Subclass&\multicolumn{1}{c}{Teff}&log $g$&[Fe/H]&Radial velocity    \\
		(yyyy-mm-dd)&(d)  &      &        &\multicolumn{1}{c}{(K)}& &    &(km$\cdot$s$^{-1}$) \\ \hline
		2011 Nov 14&2455880.390798&0.250407&F0&     & &  &            \\
		2011 Dec 07&2455903.249277&0.385092&F4&6407.81 $\pm$ 96.88&4.236 $\pm$ 0.155&-0.250 $\pm$ 0.102&70.63 $\pm$ 8.42 \\
		2014 Dec 18&2457010.244898&0.744818&F5&6267.75 $\pm$ 24.80&4.066 $\pm$ 0.033&-0.235 $\pm$ 0.018&54.21 $\pm$ 4.05 \\ \hline
	\end{tabular}}
	\label{tab:spect}
\end{table*}

\begin{figure}
	\centering
	\includegraphics[width=0.48\textwidth]{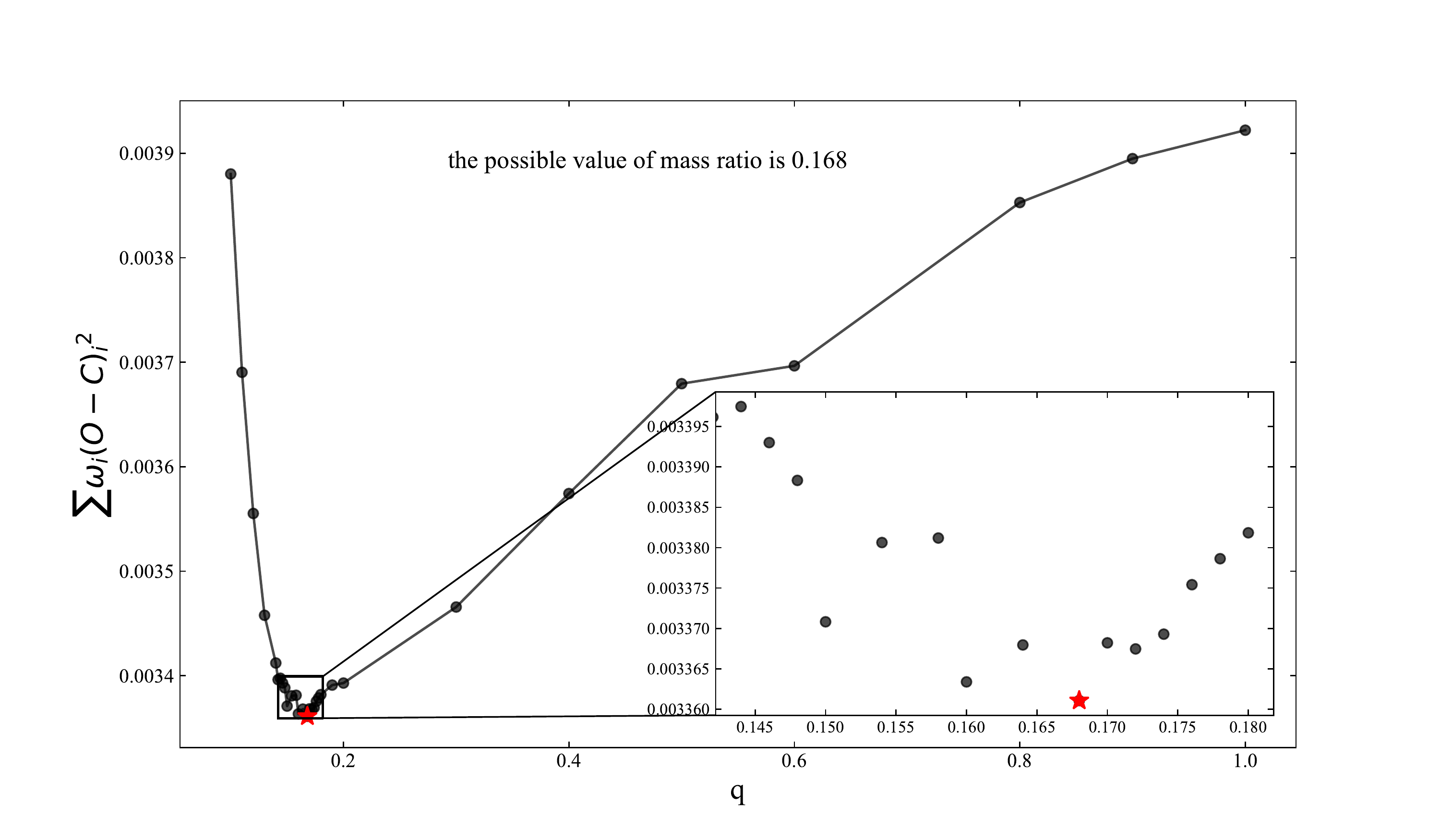}
	\caption{Relationships between sum of squares of residuals $\sum\omega_i(O - C)_i^2$ and the mass ratios $q$ for $\target$.}
	\label{fig:qsearch}
\end{figure}

The mass ratio is an important parameter for obtaining a credible photometric solution. Due to the lack of sufficient spectral observations, we used the $q$-search method to determine the mass ratio. We searched for the mass ratio of the two components between 0 and 1, i.e. primary component with more massive mass. The results are shown in Figure \ref{fig:qsearch}, where the initial value of the mass ratio $q$ entered into the W-D code as an adjustment parameter is 0.168, and this value corresponds to the smallest residual $\Sigma\omega_i(O-C)^{2}_i$. During the search for the best mass ratio, we found that the massive component has a lower temperature, which means that it is a W-type W UMa binary. Meanwhile, these two components always fill their Roche lobes. We decided to use the W-D program in mode 3 (contact state). In fact, the effective temperature of the binary star differs somewhat from the temperature of its components, and we calculate the surface temperature of each component using the following equation \citep{2013AJ....146..157C}.
\begin{align}
	&T_{1}=\left(\left(\left(1+k^{2}\right) T_{\mathrm{eff}}^{4}\right) /\left(1+k^{2}\left(T_{2} / T_{1}\right)^{4}\right)\right)^{0.25}, \\
	&T_{2}=T_{1}\left(T_{2} / T_{1}\right).
\end{align}
The $k$ in the equation is the radius ratio ($r_2/r_1$), $T_{eff}$ is the effective temperature of the binary, and $T_{2} / T_{1}$ is the temperature ratio of the two components in combination with the preliminary fitting results of the W-D program. After several iterations, the main component temperature tends to a definite value.
During the solution process, the orbital inclination $i$, the average temperature of the secondary component $T_2$, the monochromatic luminosity of the primary component $L_1$ in the $BVRI$ bands, and the dimensionless potential $\Omega_1 = \Omega_2$ of the two components are adjustable parameters.

\begin{figure}
	\begin{center}
		\includegraphics[width=0.49\textwidth]{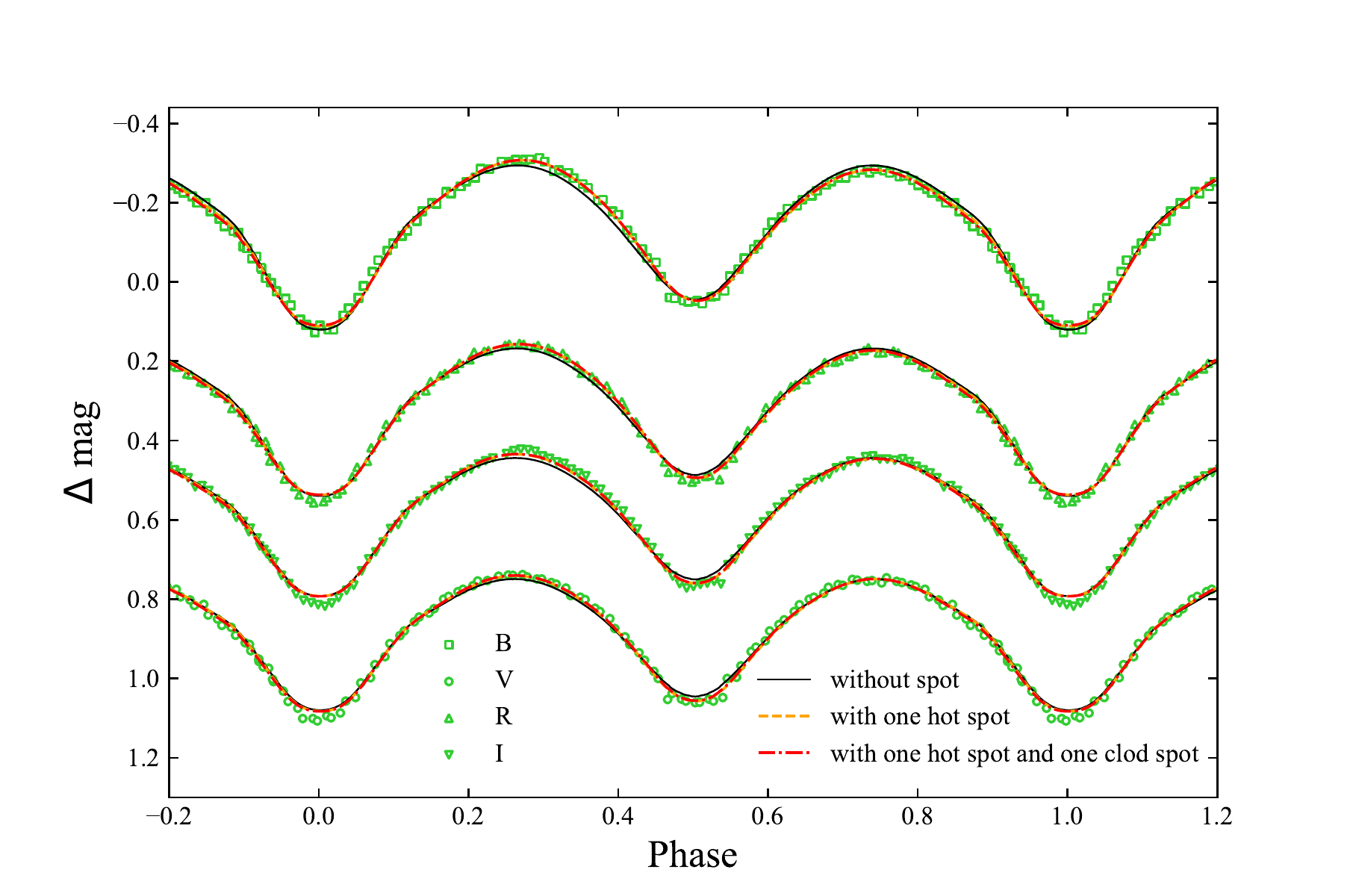}
		\caption{The photometric precisions of $\target$ in $BVRI$ band.}	
		\label{fig:wdfit}
	\end{center}
\end{figure}

\begin{figure}
	\begin{center}
		\includegraphics[width=0.49\textwidth]{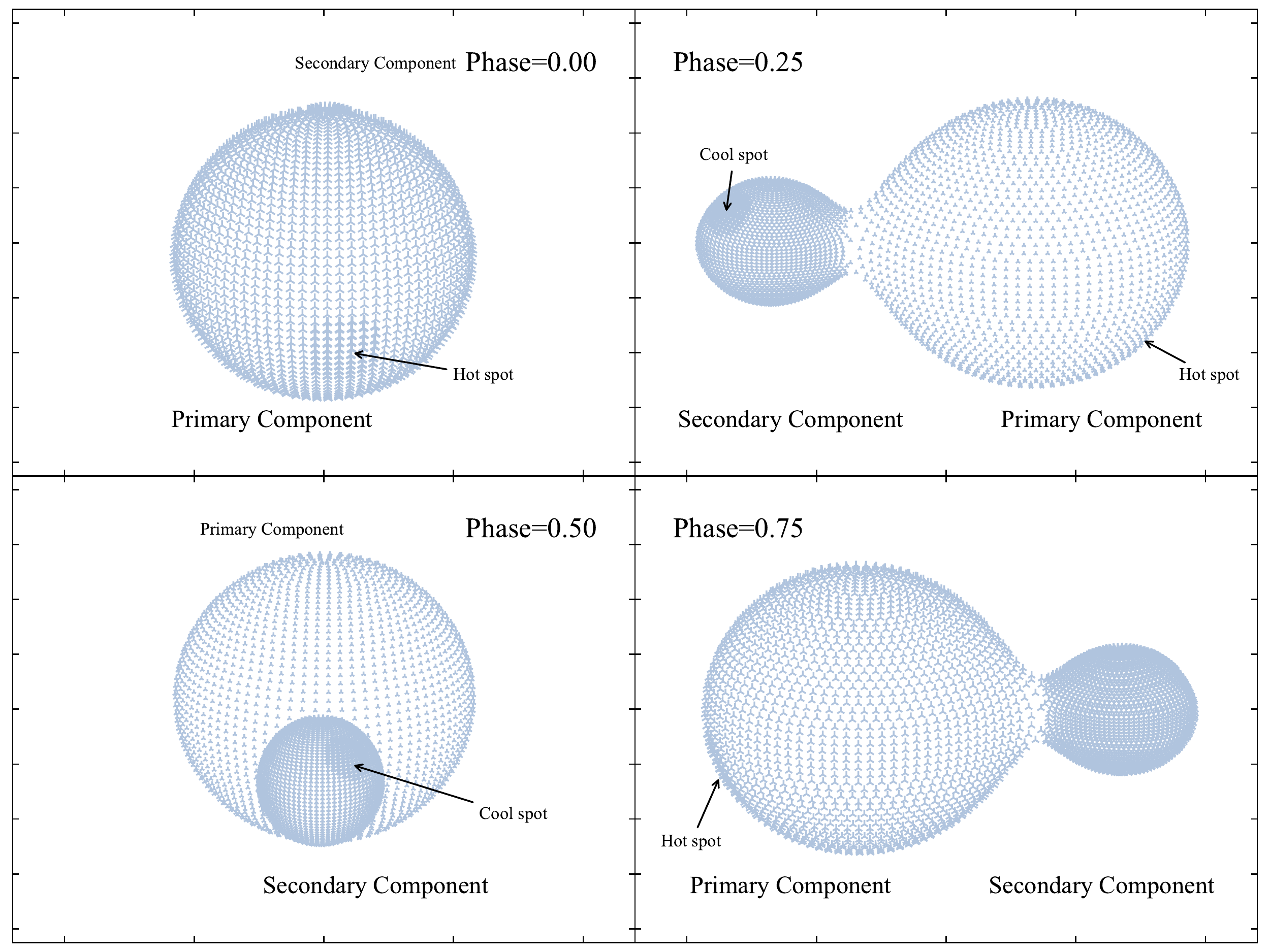}
		\caption{Geometrical structure of $\target$ with one hot spot on the primary and one cool spot on the secondary at phases of 0.00, 0.25, 0.50, and 0.75.}	
		\label{fig:model}
	\end{center}
\end{figure}

We found that the light curves are asymmetric \citep[O'Connell effect,][]{1951PRCO....2...85O} and explain this phenomenon by introducing a starspot model. The photometric solutions are listed in Table \ref{tab:elements}. The corresponding theoretical light curves that fit the observations better are displayed in Figure \ref{fig:wdfit}. From the results in Figure \ref{fig:wdfit}, introducing a hot spot in the primary component provides better convergence results, and if another cold spot is added to the secondary component, a better solution with a fill-out factor of $f$ = 50.1\% and a mass ratio of $q$ = 0.1501 was obtained, as confirmed by the residual values in Table \ref{tab:elements}. The error in the filling factor of the binary is larger at low mass ratios, because the surface potential through the inner Lagrangian point (L$_1$) and the outer Lagrangian point (L$_2$) are numerically very close. The geometrical structures of $\target$ at phases of 0.00, 0.25, 0.50, and 0.75 are plotted in Figure \ref{fig:model}.

\begin{table}[htp]
	\tiny
	\centering
	\caption{Photometric Elements of $\target$.}
	\begin{tabular}{lccc}
		\hline\hline
		Parameters&\multicolumn{1}{c}{Without spots}&\multicolumn{1}{c}{With one spot}&\multicolumn{1}{c}{With two spots} \\ \hline
		$ g_1=g_2 $&0.32&0.32&0.32                                    \\
		$ A_1=A_2 $&0.5&0.5&0.5                                     \\
		$ \Omega_{in} $ &2.07094 & 2.10336&2.10336                               \\
		$ \Omega_{out} $ &1.98144 & 2.00648 &2.00648                                 \\
		$ q (M_2/M_1) $&0.1383 ($ \pm $ 0.0026)&0.1501 ($ \pm $ 0.0016)&0.1501 ($ \pm $ 0.0013)          \\
		$ T_1 (K) $&6130 K&6130 K&6130 K                                     \\
		$ T_2 (K) $&6569 ($ \pm$ 20) K&6359 ($ \pm$ 9) K&6254 ($ \pm$ 10) K        \\
		$ i^\circ $&71.244 ($ \pm $ 0.603)&72.522 ($ \pm $ 0.345)&72.586 ($ \pm $ 0.345)         \\
		$ \Omega_{1}=\Omega_{2} $&2.02343 (0.00822)&2.05270 (0.00519)&2.05473 (0.00435)         \\
		$ L_{1B}/L_B $&0.7859 ($ \pm$ 0.0067)&0.8049 ($ \pm$ 0.0031)&0.8202 ($ \pm$ 0.0029)         \\
		$ L_{1V}/L_V $&0.8028 ($ \pm$ 0.0057)&0.8137 ($ \pm$ 0.0028)&0.8249 ($ \pm$ 0.0025)         \\
		$ L_{1R}/L_{R} $&0.8121 ($ \pm$ 0.0052)&0.8185 ($ \pm$ 0.0026)&0.8275 ($ \pm$ 0.0024)         \\
		$ L_{1I}/L_{I} $&0.8189 ($ \pm$ 0.0049)&0.8221 ($ \pm$ 0.0026)&0.8294 ($ \pm$ 0.0023)         \\
		$ r_1 (pole) $&0.5269 ($ \pm$ 0.0017)&0.5209 ($ \pm$ 0.0011)&0.5200  ($ \pm$ 0.0009 )       \\
		$ r_1 (side) $&0.5864 ($ \pm$ 0.0026)&0.5778 ($ \pm$ 0.0016)&0.5763  ($ \pm$ 0.0014 )       \\
		$ r_1 (back) $&0.6103 ($ \pm$ 0.0027)&0.6025 ($ \pm$ 0.0017)&0.6008  ($ \pm$ 0.0015 )       \\
		$ r_2 (pole) $&0.2225 ($ \pm$ 0.0088)&0.2300 ($ \pm$ 0.0049)&0.2300  ($ \pm$ 0.0040 )       \\
		$ r_2 (side) $&0.2333 ($ \pm$ 0.0108)&0.2415 ($ \pm$ 0.0061)&0.2414  ($ \pm$ 0.0049 )       \\
		$ r_2 (back) $&0.2831 ($ \pm$ 0.0288)&0.2937 ($ \pm$ 0.0164)&0.2933  ($ \pm$ 0.0131 )       \\
		$ f $&53.1 \% ($ \pm$ 9.2 \%)&52.3 \% ($ \pm$ 5.4 \%)&50.2 \% ($ \pm$ 4.5 \%)         \\ \hline
		Spot 1&&Primary &Primary          \\
		$ \theta (radian) $&&0.9291 ($ \pm$ 0.0534)&0.9291 ($ \pm$ 0.0534)           \\
		$ \phi (radian) $&&3.7145 ($ \pm$ 0.0513)&3.7145 ($ \pm$ 0.0513)           \\
		$ r (radian) $&&0.3290 ($ \pm$ 0.0092)&0.3290 ($ \pm$ 0.0092)           \\
		$ T_f (T_d/T_0) $&&1.1959 ($ \pm$ 0.0083)&1.1959 ($ \pm$ 0.0083)           \\   \hline
		Spot 2&&&Secondary        \\
		$ \theta (radian) $&&&1.9256 ($ \pm$ 0.0992)             \\
		$ \phi (radian) $&&&3.3127 ($ \pm$ 0.0518)             \\
		$ r (radian) $&&&0.2318 ($ \pm$ 0.0211)             \\
		$ T_f (T_d/T_0) $&&&0.9058 ($ \pm$ 0.0185)             \\ \hline
		$ \Sigma W (O-C) ^{2} $&0.00324&0.00203&0.00194                                   \\ \hline
	\end{tabular}
	\label{tab:elements}
\end{table}

\section{Discussion and Conclusion}

Although $\target$ had been monitored by many sky surveys, it was neglected for further study. The photometric solutions of this eclipsing binary $\target$ is presented for the first time. Using the W-D program, we analyzed one set of complete multiple color light curves of target. Our photometric solutions suggest that $\target$ is a DLMR binary system with a high degree of overcontact ($f$ = 50.1\%) and a derived mass ratio (q = 0.1501). The asymmetry of the light curve is explained by a hot spot on the primary and a cold spot on the secondary component. $\target$ is a W-type W UMa system. According to the photometric results of the three groups of component temperatures, through Table 5 of \cite{2013ApJS..208....9P}, we could conclude that the spectral type of the primary component is F8, and the spectral type of the secondary component ranges from F5 to F7. The spectral type of these daughter stars differs slightly from the LAMOST results for the spectral type of this binary system in Table \ref{tab:spect}, which is due to the different spectral type and temperature correspondence criteria.

\subsection{Absolute parameters}

Accurate absolute parameters are obtained for binary star systems with radial velocity curve. In the absence of RV curve, we could use an estimation method to obtain the absolute parameters of the binary system. Since entering the Gaia era, we are able to obtain more accurate distance information. Based on the distance and photometric observations, we are able to obtain the absolute bolometric magnitudes of the components after extinction and reddening corrections. We adopted a similar approach, assuming that the massive component of W UMa is the main sequence star \citep{2005ApJ...629.1055Y}. Firstly, the luminosity of the primary star is determined by these equations: (a) ${m}_{V_i}-{m}_{V _{\max} }=-2.5 \log \left({L}_{{i}} / {L}\right)$, (b) ${M}_{V_{i}}={m}_{V_{i}}-5 \log {D}+5-{A}_{V}$, (c) ${M}_{bol _i}={M}_{V_i}+{BC}_{V_i}$, (d) $ {M} _ {bol 1, 2} = 4.73\,+\,2.5 \log \left({L} _ {1, 2} /  {L} _ {\odot}\right) $ \citep{2010AJ....140.1158T}. In our work, the interstellar extinction coefficient $A_{v_{S \& F}}$=0.1701$\pm$0.0022 (mag) \citep{2011ApJ...737..103S} from IRSA database\footnote{\url{https://irsa.ipac.caltech.edu/applications/DUST/}}, the bolometric corrections $BC_v$=-0.042 (mag) from \citet{2013ApJS..208....9P}, distances obtained by \citet{2022yCat.1355....0G}, and the maximum visual magnitude $m_{V_{\max}}$ obtained by fitting ASAS-SN data. \citet{2022MNRAS.512.1244C} provided a mass-luminosity relationship $\log{L}=\log{b}+a\log{M}$ ($b= 0.63 \pm 0.04$ and $a= 4.8 \pm 0.2$). The equations ${M}_{1}+{M}_{2}=0.0134 {a}^{3} / {P}^{2}$ and $R_{i}=a r_{i}$ were used. The calculated results are shown in Table \ref{tab:abs}. Through the mass luminosity relationship, $M_1$=  1.331(1) M$_\odot$ and $M_2$= 0.200(2) M$_\odot$ were obtained.

\begin{table}
	\caption{Absolute Parameters of $\target$.}
	\centering
	\begin{tabular}{lcc}
		\hline\hline
		Parameters           & Value & Error \\
		\hline%
		$D$(pc)           &     1179.078   &         25.544  \\
		$m_{Vmax}$(mag)   &     14.104     &       0.025     \\
		$m_{V_1}$(mag)    &     14.313     &       0.028     \\
		$M_{V_1}$(mag)    &     3.785      &      0.078      \\
		$M_{bol_1}$(mag)  &     3.743      &      0.078      \\
		$L_1$(L$_\odot$)  &     2.482      &      0.177      \\
		$L_2$(L$_\odot$)  &     0.527      &      0.047      \\
		$M_1$(R$_\odot$)  &     1.331      &      0.001      \\
		$M_2$(R$_\odot$)  &     0.200      &      0.002      \\
		$a$(R$_\odot$)    &     4.800      &      0.200      \\
		$R_1$(M$_\odot$)  &     1.470      &      0.004      \\
		$R_2$(M$_\odot$)  &     0.660      &      0.019      \\
		\hline
	\end{tabular}
	\label{tab:abs}
\end{table}

\subsection{Evolutionary States}

\begin{figure}[!htbp]
	\begin{center}
		\includegraphics[width=0.48\textwidth]{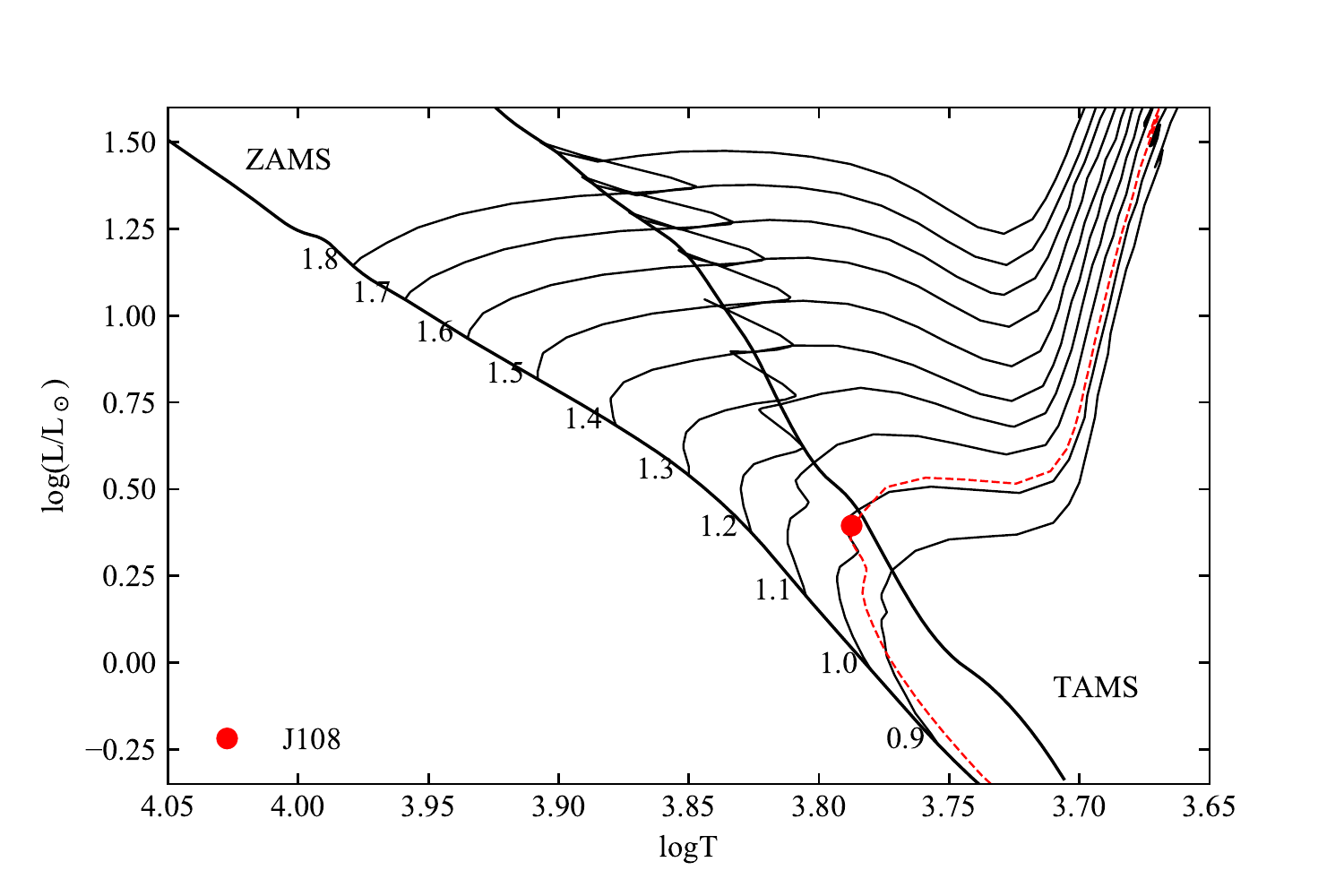}
		\caption{Primary component for $\target$ in the H-R diagram. the zero-age main sequence (ZAMS), the terminal-age main sequence (TAMS), and the evolutionary tracks of each initial mass (from 0.9 M$_\odot$ to 1.8 M$_\odot$) and isochrones around the primary component are taken from \cite{2000A&AS..141..371G} with metal abundance Z = 0.008. The red dotted line are isochrone line.}	
		\label{fig:evolution}
	\end{center}
\end{figure}

Based on the absolute parameters, the primary component is plotted in a temperature-luminosity diagram, which is shown in Figure \ref{fig:evolution}. Combined with the [Fe/H] information provided by the LAMOST spectra in Table \ref{tab:spect}, corresponding to a Z of about 0.008$\pm$0.002, the tracks of stellar evolution taken from \cite{2000A&AS..141..371G} with metal abundance Z = 0.008 were chosen. The primary is located between the zero-age main sequence (ZAMS) and the terminal-age main sequence (TAMS) lines, which suggests that the primary component is an evolved MS star. Based on the evolutionary tracks of the single stars, we estimated the possible evolutionary age of $\target$ to be $\thicksim$7.94 Gyr.

\begin{figure*}[!htbp]
	\centering
	\includegraphics[width=0.48\textwidth]{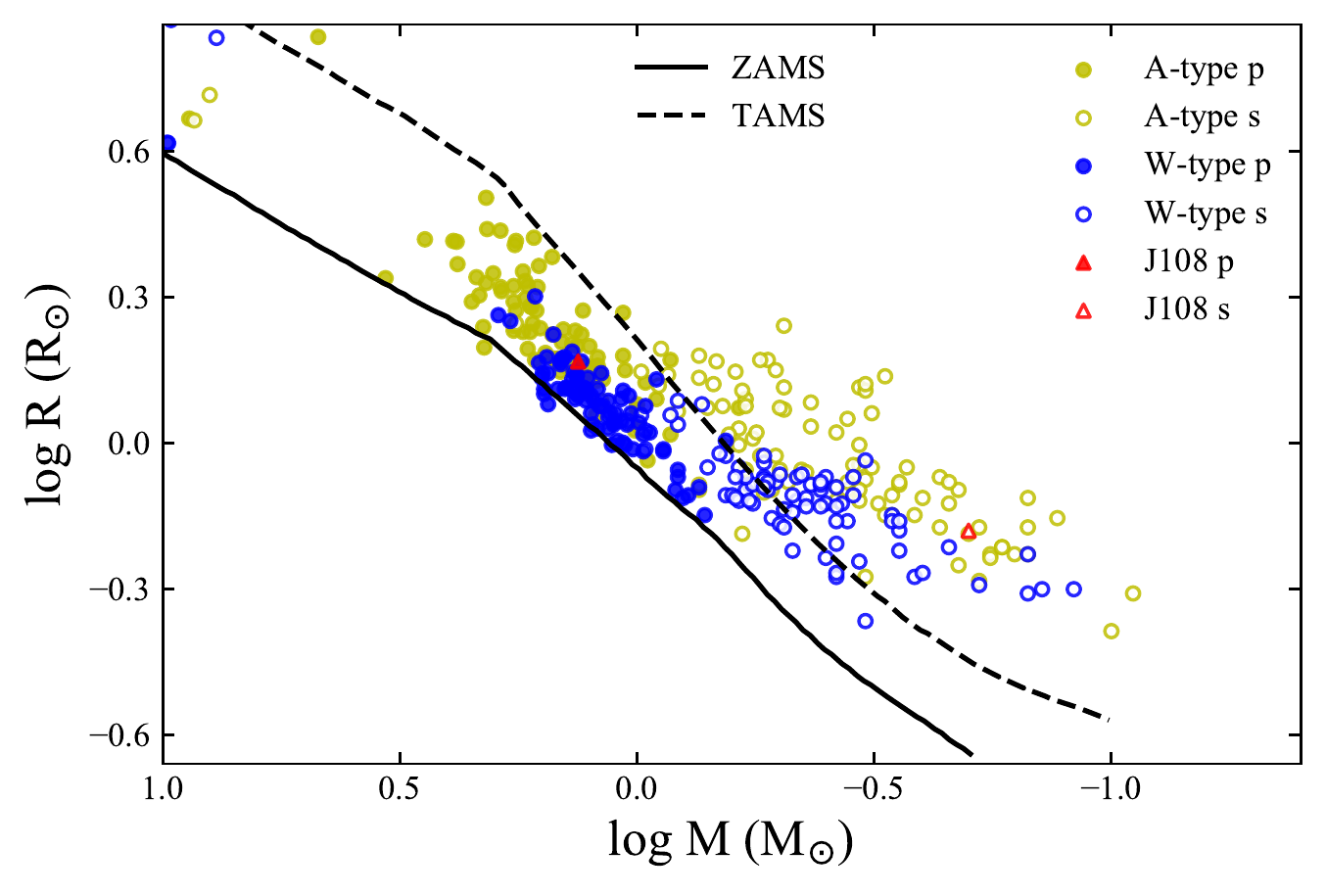}
	\includegraphics[width=0.48\textwidth]{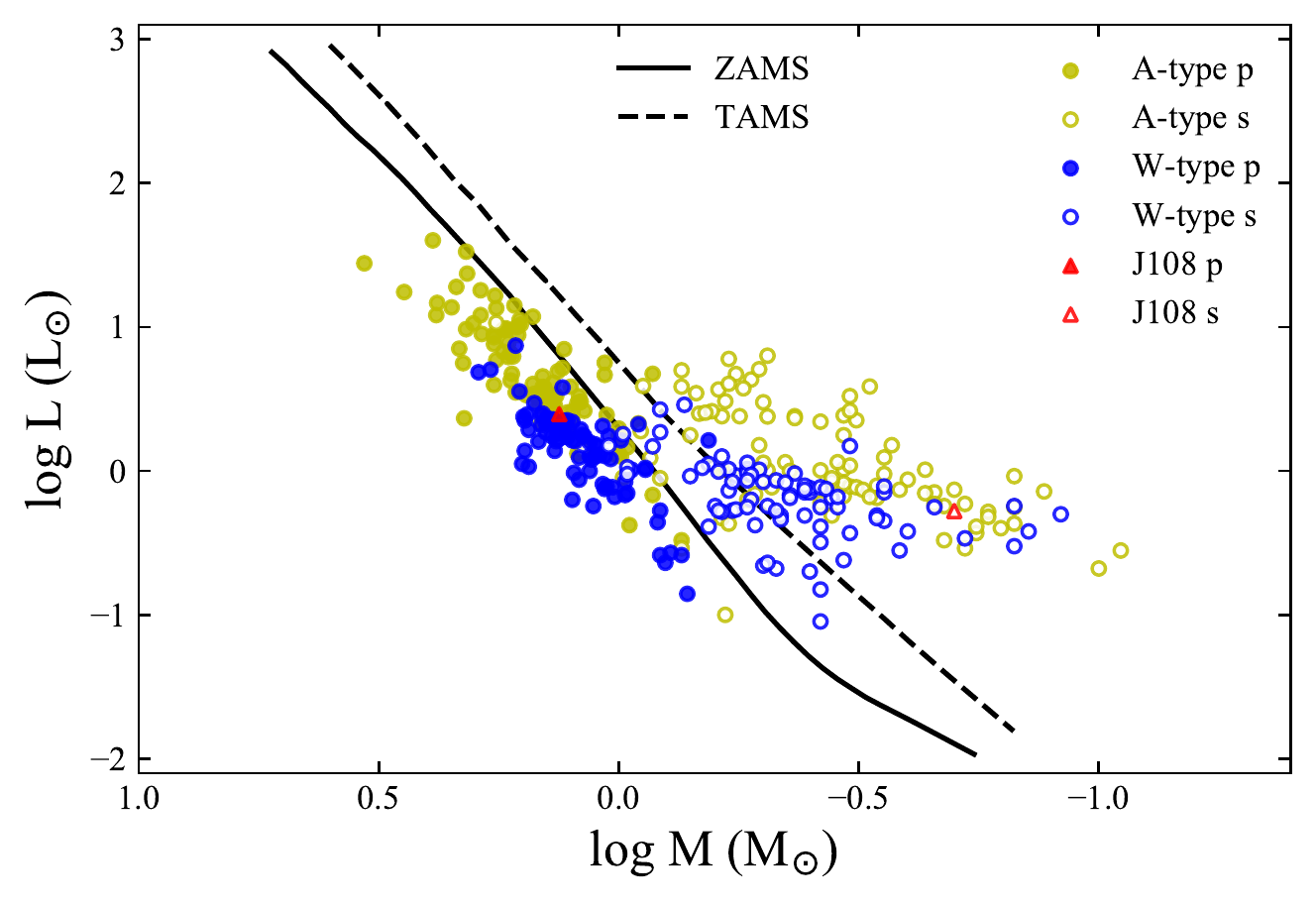}
	\caption{The mass-radius diagram (left) and mass-luminosity diagram (right). The solid line and the dashed line, which represent the ZAMS and TAMS respectively, are constructed by the binary system evolution code provided by \citet{2002MNRAS.329..897H}. Red regular triangles represent our own two targets. The A-type and the W-type contact binaries from \citet{2021AJ....162...13L} are displayed as yellow circles and blue circles, respectively. }
	\label{fig:mass-}
\end{figure*}

The mass-radius (M-R) and mass-luminosity (M-L) relations were used to understand the evolutionary state of our target binary. $\target$ with from Table 7 of \citet{2021AJ....162...13L} for 94 A-type and 85 W-type binaries are plotted in Figure \ref{fig:mass-}. In the figure, the target, like most binary systems, has the higher mass component as the evolved main-sequence star, while the lower mass component of the target has left the main-sequence phase. This could be a mass transfer leading to a mass ratio reversal \citep{1988ASIC..241..345G}, meaning that the current low-mass component is the initial high-mass component of the binary system, and this component underwent a rapid mass transfer process as it left the main-sequence phase.

Determining the initial masses of the two components is useful to study the evolution of contact binaries. Since binaries are influenced by their companions during their evolution, the method of luminosity excess \citep{2013MNRAS.430.2029Y} was used to calculate the initial mass of the initial massive component (currently the lower mass component) before the mass transfer process occurred.The initial mass of the secondary $M_{2\mathrm{i}}=1.83\pm 0.04$ M$_\odot$ was obtained using the following equation:
\begin{align}
	M_{2 \mathrm{i}}&= M_{2}+2.50(\delta M-0.07)^{0.64}, \\ \nonumber
	\Delta M_{2 \mathrm{i}}&=\Delta M_{2}+\frac{1.59}{(\delta M-0.07)^{0.36}}\left(\frac{0.22 \Delta L_{2}}{L_{2}^{0.76}}+\Delta M_{2}\right),
\end{align}
where $\delta M=M_{L}-M_{2}$, $ M_{2}$ is the current mass of the secondary component, $M_{L}=\left(L_{2} / 1.49\right)^{1 / 4.216}$, $\Delta$ denote the errors of the corresponding quantity. Based on the possible initial mass ratios, the initial mass range of the primary star was determined ($M_{1\mathrm{i}}$ from 0.67 to 1.21 M$_odot$). The initial mass of 0.985 M$_\odot$ obtained from Figure \ref{fig:evolution} is in this initial mass range, assuming that the primary star is a main sequence single star. The thermal timescale of the more massive star is shorter. When the mass difference between the two components of a binary system is relatively large (about 2.5 times), the evolution of the massive component is faster \citep{2022A&A...659A..98S}. The outer layers of the more massive component expand and break through its Roche radius. The less massive component is subject to unstable dynamical mass transfer and gains mass. The previous time scale required for the thermal adjustment of the smaller mass component is longer than the time scale required for the mass loss of the larger mass component, so the less massive star will expand. When both components expand beyond the Roche critical volume, the binary star evolves into a system with a common envelope.\citep{2010NewAR..54...65T,2013A&ARv..21...59I} It can be seen that the initial massive component lost a large amount of mass (about 1.6 M$_\odot$) when it left the main sequence phase, part of which was transferred to the present primary star (about from 0.1 to 0.66 M$_\odot$), and another more mass was lost from the binary system (about from 1 to 1.5 M$_\odot$), which will cause a decrease in the orbital angular momentum of the binary system. The process could be called as the roche lobe overflow \citep{2002MNRAS.329..897H,2022ApJ...937L..42H}, during which the initial less massive component gains a small fraction of the mass lost by the massive component, and more mass is spilled from the outer Lagrangian points in order to form short-period binary systems, a theory corroborated by our observations.

The equation $\frac{J_{\text {spin }}}{J_{\text {orb }}}=\frac{1+q}{q}\left[\left(k_{1} r_{1}\right)^{2}+\left(k_{2} r_{2}\right)^{2} q\right]$ was given by \citet{2015AJ....150...69Y} to calculate the ratio of the spin angular momentum $J_{\text{spin}}$ to the orbital angular momentum $J_{\text {orb }}$, which reflects the stability of the binary evolution. In this work, $k^2_i$ is set as 0.06 \citep{2006MNRAS.369.2001L} and $\frac{J_{\text {spin }}}{J_{\text {orb }}}=0.151$ was obtained, $J_{\text{spin}} / J_{\text{orb}}<1/3$ indicate that $\target$ are currently in a stable evolutionary state.

This paper presents the first detailed photometric analysis of ATO J108.6991+27.8306 using multimetric photometric data from NOWT observations. This is a neglected, deep ($f$ = 50.1 \%), low mass ratio ($q$ = 0.1501) W-type W UMa binary system, which implies that it is at a late evolutionary stage of contact binary systems. We have derived the initial masses of the two components and inferred an evolutionary age of the system of about 7.94 Gyr.

The minima of light times we have collected so far are not enough to support our work on long-term orbital period analysis, so more observations are necessary and the variation of the orbital period needs to be further investigated. The continuous loss of orbital angular momentum is visualized by the decreasing orbital period of the system, which may lead to an increase of the filling factor and a decrease of the inter-binary distance, which will evolve into a rapidly rotating star. Conversely, if the orbital period of the target is continuously increasing, then ATO J108.6991+27.8306 will be more valuable as an idiosyncratic target for research, and more detailed data are needed to support the construction of the theory.

\begin{acknowledgements}
	This work received the generous support of the National Natural Science Foundation of China under grants U2031204. We gratefully acknowledge the science research grants from the China Manned Space Project with NO.CMSCSST-2021-A08. New CCD photometric observations of the system were obtained with the 1.0 m telescope (NOWT) at Xinjiang Observatory.
\end{acknowledgements}

\bibliographystyle{raa}
\bibliography{bibtex}

\end{document}